\documentclass[aip,jcp,amsmath,amssymb,floatfix,reprint,citeautoscript]{revtex4-2}

\usepackage[utf8]{inputenc}
\usepackage{graphicx} %
\usepackage{wrapfig} %
\usepackage{ae} %
\usepackage{booktabs} %
\usepackage{silence}
\usepackage{float}
\usepackage{gensymb} %
\usepackage{tikz} %
\usepackage{multirow}
\usepackage{chemscheme} %
\usepackage[colorlinks,allcolors=black,citecolor=blue,urlcolor=blue]{hyperref}

\begin{document}

\title{How Alkali Metal Cations Affect the Structure and Reactivity of the Hydrated Dielectron}

\author{Tatiana Nemirovich}
\affiliation{Institute of Organic Chemistry and Biochemistry of the Czech Academy of Sciences, Flemingovo nám. 2, 166 10 Prague 6, Czech Republic}

\author{Pavel Jungwirth}
\affiliation{Institute of Organic Chemistry and Biochemistry of the Czech Academy of Sciences, Flemingovo nám. 2, 166 10 Prague 6, Czech Republic}

\author{Ondrej Marsalek}
\email{ondrej.marsalek@matfyz.cuni.cz}
\affiliation{Charles University, Faculty of Mathematics and Physics, Ke Karlovu 3, 121 16 Prague 2, Czech Republic}

\date{\today}

\begin{abstract}

Hydrated electrons of opposite spins pair to form dielectrons at sufficiently high concentrations that can be achieved by dissolution of alkali metals in water.
While experimental investigations of these systems are challenging due to their vigorous, even explosive, reactivity, simulations open the possibility to characterize the structure and reactivity of hydrated dielectrons and the effects of alkali cations thereon.
Here, we present \textit{ab initio} molecular dynamics simulations of a hydrated dielectron without or with explicit Li$^+$ or Cs$^+$ counterions.
While the overall solvation structure is preserved in all these systems, the presence of cations has a distinct effect of increasing the dielectron gyration radius by about 10\% and forming cation-specific cation--dielectron arrangements.
Moreover, analysis of water bond lengths reveals in all studied systems a substantial elongation of first-shell O--H bonds oriented toward the dielectron, providing a structural explanation for vibrational red-shifts observed in resonance Raman measurements.
Finally, at the 10 ps simulation timescale, rare reactive events were observed, albeit only for the system without metal cations, where hydride intermediates stable on picosecond timescales were identified. 
These observations also suggest that on the investigated timescales, metal cations may suppress hydrated dielectron reactivity.

\end{abstract}

\maketitle

\section{Introduction}

Dissolution of alkali metals in water leads to the formation of hydrated electrons, i.e., excess electrons localized in liquid water and stabilized by the surrounding solvent environment.
These species are among the strongest known reducing agents in aqueous solution~\cite{hart1969research, rybkin2020mechanism} and participate in a wide range of chemical and biological processes~\cite{herbert2017hydrated}, including radiation-induced DNA damage\cite{garrett2004role, kumar2019reaction}.
Hydrated electrons exhibit characteristic optical absorption spectra~\cite{hart1962absorption} and unique solvation structures that have made them a long-standing subject of both experimental~\cite{hart1962absorption, bartels2001moment, tauber2003structure, bartels2005pulse, coe2008photoelectron, Coulomb_explosion,  mason2021spectroscopic} and theoretical investigation~\cite{herbert2017hydrated, rybkin2020mechanism, reactions_gao, Schwartz_contact, reaction_borrelli, kar2025nature, Diaz2026/10.1038/s41467-026-70045-7}.
Two single excess electrons can become spin-paired, forming a hydrated dielectron~\cite{guardado2026ab}, in which both electrons occupy a common solvent cavity in an overall singlet state, becoming increasingly relevant at elevated electron concentrations~\cite{vitek2025dynamics}.

Compared to the extensively studied hydrated electron, the hydrated dielectron remains significantly less explored, even though it has long been proposed as a key reactant in hydrogen evolution and in the reduction chemistry of alkali metals in water~\cite{reaction_bartels, reaction1, reaction_cluster, reaction_borrelli, reactions_gao}.
The chemistry of excess electrons in aqueous solution is complex and includes a wide range of competing reaction pathways~\cite{garrett2004role}.
While the hydrated electron readily reacts with protons to form hydrogen atoms, its reaction with water molecules is many orders of magnitude slower than that of the hydrated dielectron~\cite{garrett2004role}.

The currently accepted mechanism of hydrogen evolution from hydrated dielectrons involves two sequential proton-transfer steps~\cite{reaction_bartels, reaction_borrelli, reactions_gao}.
In the first step, the dielectron abstracts a proton from a neighboring water molecule, yielding a hydride anion and hydroxide ion.
The hydride subsequently accepts a second proton from another water molecule, producing molecular hydrogen and a second hydroxide ion:
\begin{align}
e_{2,\mathrm{aq}}^{2-} + \mathrm{H_2O}
&\rightarrow
\mathrm{H^-} + \mathrm{OH^-}
\\
\mathrm{H^-} + \mathrm{H_2O}
&\rightarrow
\mathrm{H_2} + \mathrm{OH^-}
\end{align}
The existence of a hydride intermediate has been proposed for several decades and is supported by experimental observations of hydrated hydride species in irradiated aqueous systems~\cite{kimmel1994low, bernas1977existence}.
However, direct characterization of the reaction pathway remains challenging due to the high reactivities and short lifetimes of both the dielectron precursor and the hydride intermediate.

Higher transient concentrations of hydrated electrons are most easily achieved by dissolving alkali metals in water.
Despite the obvious presence of metal cations in these solutions, computational studies explicitly including both excess electrons and metal cations remain relatively limited.
Most such studies focus on the reaction of neutral metal atoms with water cluster and therefore probe highly non-equilibrium early stages of electron transfer from the metal rather than equilibrated aqueous electron--cation systems~\cite{mundy2000microsolvation, mercuri2001formation, cwiklik_sodium, Coulomb_explosion}.
In addition, finite water clusters may not adequately capture the solvation structure and dynamics of bulk liquid water.
Only recently has attention turned to hydrated electron--cation contact pairs, where it was shown that ion pairing can influence electron localization, hydration structure, and spectroscopic signatures~\cite{Schwartz_contact}.
However, none of the existing ab initio molecular dynamics (AIMD) studies of hydrated dielectrons in bulk water explicitly included metal cations, employing thus non-electroneutral simulation cells~\cite{reaction_cluster,reaction_borrelli,reactions_gao}.
In contrast, realistic alkali-metal aqueous solutions are electroneutral and necessarily contain both excess electrons and counter cations.
Including metal cations, therefore, not only provides a more physically realistic description of the system, but is also essential for capturing cation-dependent behavior.
Indeed, aqueous alkali-metal solutions are known to exhibit pronounced cation-dependent optical properties~\cite{vitek2025dynamics}.
Together, these observations suggest that metal cations are not merely spectator ions, but may significantly influence the structure, stability, and reactivity of hydrated dielectrons.

In the present work, we address this knowledge gap by introducing alkali metal cations into AIMD simulations of hydrated dielectrons in bulk water.
Specifically, we compare systems without cations with systems containing Li$^+$ and Cs$^+$ ions, representing the limiting cases of chemical hardness and softness.
By analysis of radial distribution functions, gyration radii, and shape descriptors we describe how the presence of cations affects the structure and properties of hydrated dielectrons.
In addition, we examine dielectron--cation interactions, the differences between Li$^+$ and Cs$^+$ spatial arrangements around the dielectron and their influence on the localization and shape of the excess electron.
Finally, we compare reactivities of systems with and without cations and, in cases where reactive events are observed, characterize the resulting intermediates and their structural and solvation properties.

\section{Methods}

\subsection{Ab initio molecular dynamics}

The AIMD simulations were performed using the CP2K~9.1 software package~\cite{hutter2014cp2k} with the Quickstep module~\cite{vandevondele2005quickstep, quickstep}.
All studied systems systems were simulated under periodic boundary conditions.

A hybrid density functional with dispersion correction, revPBE38-D3~\cite{perdew1996generalized, adamo1999toward, zhang1998comment, goerigk2011thorough}, was employed together with GTH pseudopotentials for core electrons~\cite{GTHpseudopotentials} and the TZV2P-GTH basis set for valence electrons~\cite{vandevondele2005quickstep}.
The chosen functional incorporates a relatively high proportion (3/8 = 37.5\%) of exact exchange, which has been found to be essential for describing systems susceptible to metal--insulator transitions \cite{pavlak2024electronic} and for a more reliable description of solvated electrons~\cite{ambrosio2017electronic, baranyi2020ab, carter2023birth, wilhelm, herbert, borrelli2025choice}.
The electronic density was represented using the GPW scheme with a plane-wave cutoff of 400~Ry and the self-consistent field (SCF) convergence threshold was set to $5 \times 10^{-6}$.
The auxiliary density matrix method (ADMM) was applied using the cpFIT3 basis set~\cite{guidon2010auxiliary} (cpFIT11 for Cs atoms) within the restricted Kohn--Sham framework.
The dispersion correction (D3) was disabled for metal cations due its known unphysical behavior with alkali metal cations~\cite{kostal2023common}.

Classical equations of motion were integrated with a time step of 0.5~fs within the canonical (NVT) ensemble.
Forces were obtained from an electronic structure calculation at each step.
During production runs, the temperature was controlled at 298~K using the global CSVR thermostat~\cite{CSVR} with a time constant of 500~fs.

For the systems containing cations, 10 independent trajectories were performed, each with a production time of 10~ps.
For the system without cations, a total of 30 trajectories were generated: the first 10 trajectories were propagated for 10~ps, while the remaining 20 trajectories were run for 5~ps each.

The initial structures were generated using force field molecular dynamics in the GROMACS package~\cite{gromacs}.
An iodide anion placeholder (I$^-$) was used instead of the dielectron.
In this way, an initial cavity was created that is suitable for the solvated electron due to its size and polarization of the surrounding water molecules.
For the GROMACS trajectories, the CHARMM27 force field~\cite{charmm} with the SPC/E water model~\cite{spce} was used.
Following energy minimization, a 20~ns equilibration run was performed in a canonical (NVT) ensemble.
From this trajectory, random snapshots were extracted and used as starting geometries for subsequent equilibration runs is CP2K with the I$^-$ anion present.

For the initial system preparation at the DFT level, Langevin dynamics~\cite{Langevin} at 298~K with a friction coefficient of 0.02~ps$^{-1}$ was employed.
These simulations were carried out using the revPBE~\cite{perdew1996generalized} functional, GTH pseudopotentials for core electrons~\cite{GTHpseudopotentials}, and the TZV2P-GTH basis set~\cite{vandevondele2007gaussian}, with a time step of 0.5~fs.
During this stage, the I$^-$ anion was still present as a placeholder, and dispersion corrections for the metal cations (D3) were disabled~\cite{kostal2023common}.
After 3~ps, the I$^-$ anion was removed and replaced by the dielectron (charge $-2$), which was then used in the subsequent hybrid production AIMD trajectories.

Since the dielectron is spin-paired, spin density cannot be used to characterize the distribution of the excess charge.
Instead, the spatial localization of the hydrated dielectron was analyzed using maximally localized Wannier functions (MLWFs), whose square represents the probability density of the dielectron.
From this probability density, we computed the center of the dielectron and the gyration tensor, which provides access to the radius of gyration $r_\mathrm{g}$ as a measure of the spatial extent of the excess electron.
In addition, we evaluated other shape descriptors derived from the gyration tensor, including the asphericity, acylindricity, and the relative shape anisotropy $\kappa^2$.
The parameter $\kappa^2$ ranges from 0 to 1, where 0 corresponds to a perfectly spherical distribution and values approaching 1 indicate increasing deviations from spherical symmetry.
The mathematical definitions of these quantities are well established in the literature~\cite{theodorou1985shape, marsalek2010hydrogen} and are therefore not repeated here.
Averaging the probability density over the angles in spherical coordinates centered on the dielectron yields a radial density profile of the dielectron.

To analyze the effective instantaneous O--H bond distances, it was necessary to remove the contribution of high-frequency vibrational motion.
For this purpose, a low-pass filtering procedure was applied to the time evolution of each O--H bond.
The time evolution of each O--H bond were transformed into the frequency domain using fast Fourier transform, and frequency components above 1000~cm$^{-1}$ were removed.
An inverse Fourier transform was then performed to reconstruct the filtered time series.
This procedure yields a smoothened representation of the O–H bond distances, effectively isolating slower structural evolution from the instantaneous vibrational motion.

\subsection{
Simulated systems
\label{sec:simulated-systems}
}

\begin{figure}
    \centering
    \includegraphics[width=\linewidth]{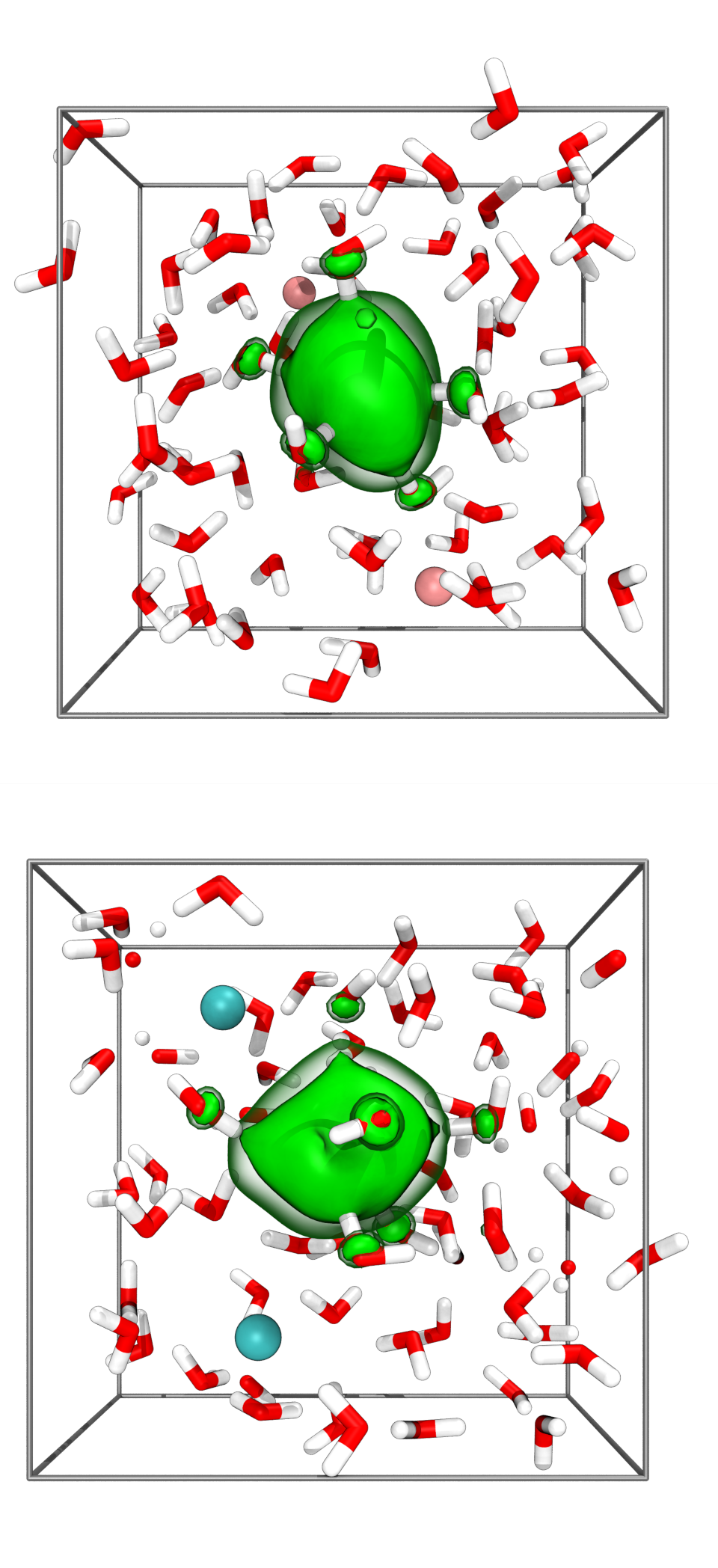}
    \caption{
    Representative snapshots of the simulated systems systems.
    Top panel: Li$^+$ system.
    Bottom panel: Cs$^+$ system.
    }
    \label{fig:snapshots}
\end{figure}

In this study, we perform bulk AIMD simulations of liquid water (64 molecules) containing a hydrated dielectron.
Three systems were considered: (i) pure water with one dielectron (net charge of the simulation box $-2$), (ii) water with one dielectron and two lithium cations (overall neutral system), and (iii) water with one dielectron and two cesium cations (again overall neutral).
The alkali metal cations were chosen to represent limiting cases of ion size and interaction strength, with Li$^+$ and Cs$^+$ corresponding to small, strongly interacting and large, weakly interacting cations, respectively.

The simulation box size was determined based on the experimental density of water and the experimental partial molar volume of the hydrated electron\cite{janik2019partial}, resulting in a cubic box with a side length of 12.53~\AA\ for the system without cations.
For the lithium-containing system, the partial molar volume of Li$^+$ in water is slightly negative and was therefore neglected as a first approximation, leading to the same box size. 
For the cesium system, the box size was based on experimental density of Cs salts solution in water~\cite{lide2004crc} and was estimated to be 12.64~\AA.

\section{Results}

Here we present the structural and dynamical properties of hydrated dielectrons obtained from the AIMD simulations for all the investigated systems.
The first 1~ps of each trajectory was treated as equilibration and excluded from further analysis.

\subsection{Structure of hydrated dielectrons}

\begin{figure*}
    \centering
    \includegraphics{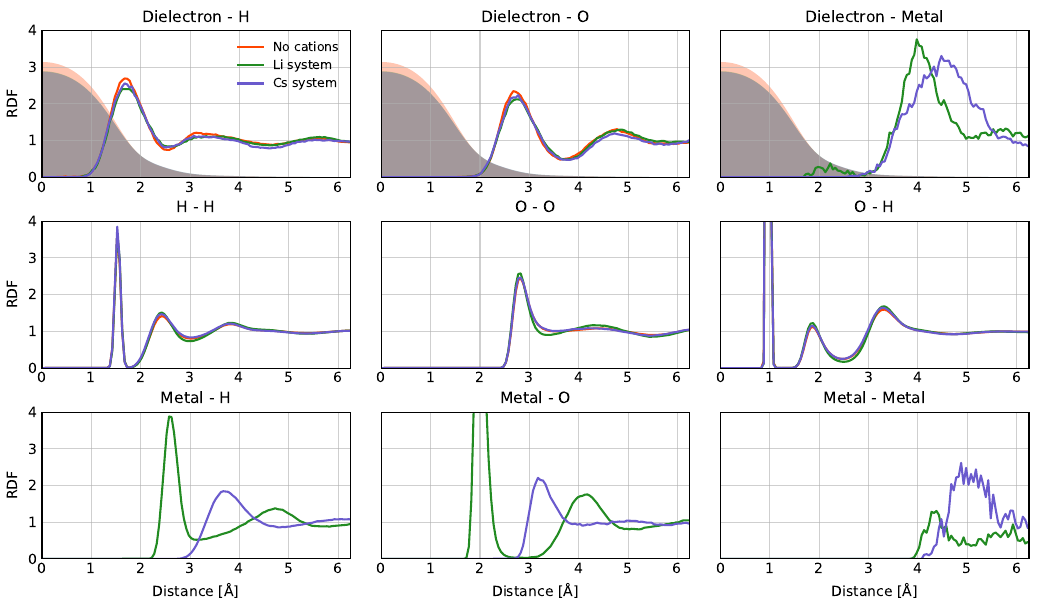}
    \caption{
    Radial distribution functions averaged over all trajectories for each system.
    For dielectron-related RDFs (top row), distances are computed with respect to the center of the MLWF.
    In the dielectron–hydrogen RDF (top left panel), radial electron density profiles obtained from the squared MLWFs are shown as filled curves.
    }
    \label{fig:rdfs-all}
\end{figure*}

Radial distribution functions (RDFs) of the dielectron were computed for all systems using the center of the MLWF corresponding to the hydrated dielectron and are shown in Figure~\ref{fig:rdfs-all}.
Separate RDFs for each system (no cations, Li$^+$ and Cs$^+$) together with integrated coordination numbers are provided in the SI (Figure~S1--S3).

The dielectron--hydrogen RDF exhibits a pronounced first peak at approximately 1.9~\AA\ in all systems.
The dielectron--oxygen RDF exhibits a pronounced first peak at 2.6~\AA, with the first solvation shell extending up to 3.6~\AA.
The absence of hydrogen or oxygen atoms within 1.9~\AA\ from the dielectron density center indicates the formation of a cavity around the electron.
These distances of the first peaks provide an estimate of the size of this cavity.
At the same time, the fact that the oxygen peak is roughly 1~\AA{} further than the hydrogen peak confirms the preferential orientation of water molecules with their O--H bonds pointing toward the center of the electron cavity.
Integration of the first solvation shell in the dielectron--oxygen and dielectron--hydrogen RDFs yields a coordination number of approximately 6--7 water molecules (See Figure~S1--S3).
Radial electron density profiles of the dielectron were also computed and are shown in each of the panels in the top row of Figure~\ref{fig:rdfs-all}.
They are very similar in the three systems, with the one with no cations being slightly higher at the center of the dielectron.
A notable feature is the substantial overlap between the radial electron density profile and the first peaks of both the dielectron--hydrogen and dielectron--oxygen RDFs.
This overlap suggests that a fraction of the dielectron density partially occupies covalent OH bonds or antibonding orbitals of the surrounding water molecules, as illustrated in Figure~\ref{fig:snapshots}.

Next, we examine the dielectron--metal RDFs (top right panel of Figure~\ref{fig:rdfs-all}).
For both metal cations, the first RDF peak appears beyond 3~\AA{}, and no overlap is observed between the dielectron radial density profile and this first RDF peak.
In the lithium-containing system, a small peak is observed around 2~\AA{}, suggesting that a small fraction of Li$^+$ ions forms contact pairs with the electron.
A more detailed inspection of our data reveals that in one case a dielectron--Li$^+$ contact pair configuration formed after 3~ps and remained stable until the end of the trajectory. 
No such contact interactions were observed for Cs$^+$.
These dielectron--cation RDFs indicate that no disproportionation or strong localization of the dielectron at the cations occurs.
That means that the system stays in a configuration with a single dielectron and two metal cations, as opposed to one metal cation and one metal anion~\cite{dye2006role, ceraso197423na}.

The water--water RDFs (H–-H, O–-O, and O–-H; middle row of Figure~\ref{fig:rdfs-all}) are nearly identical across all systems.
Minor deviations observed for the Li$^+$ system might arise from the treatment of the simulation box size, where the slightly negative partial molar volume of Li$^+$ in water was neglected, resulting in the same box size as for the system with no cations (see Section~\ref{sec:simulated-systems} for more details).
At the same time, the similarity of the total RDFs should be interpreted with caution.
These functions include contributions from all solvent molecules, including those far from the electron, where the structure is only weakly perturbed.
To resolve these effects, a more local analysis was performed by computing separate water--water RDFs for molecules closest to the dielectron (within a 4~\AA\ cutoff) and the remaining solvent.
These RDFs are shown in Figures~S4 and S5.
However, no clear or systematic differences allowing straightforward conclusions about the influence of metal cations on local water--water structure around the dielectron could be identified from the obtained data.

Overall, the RDFs are very similar across all systems, indicating that the solvation structure of the dielectron is largely unaffected by the presence of metal cations.

\begin{figure}
    \centering
    \includegraphics{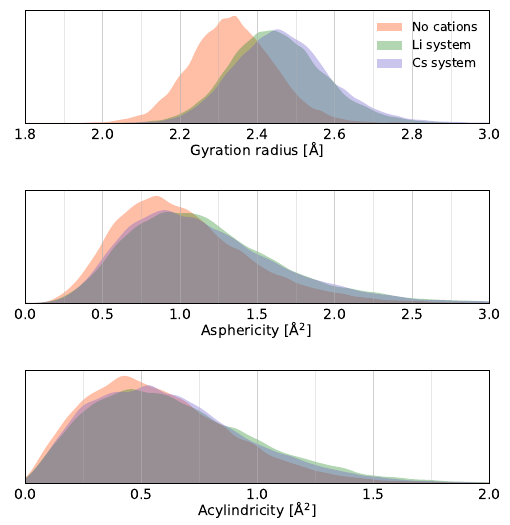}
    \caption{Distributions of the gyration radius $r_\mathrm{g}$ (top panel) and shape descriptors --- asphericity (middle panel) and acylindricity (bottom panel) for all studied systems.
}
    \label{fig:gyration-descriptors}
\end{figure}

\begin{figure*}[!tb]
    \centering
    \includegraphics{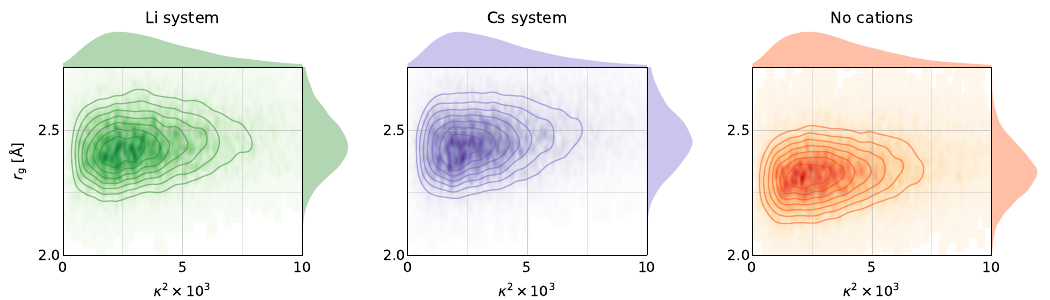}
    \caption{Two-dimensional distributions of gyration radius $r_{\mathrm{g}}$ and relative shape anisotropy $\kappa^2$ for the Li$^+$ system (top), Cs$^+$ system (middle), and the system without cations (bottom). Contour lines are smoothed using a Gaussian filter with $\sigma = 2.0$.}
    \label{fig:2D-r-k}
\end{figure*}

The distribution of gyration radii $r_\mathrm{g}$ of all the studied systems is shown in top panel of Figure~\ref{fig:gyration-descriptors}.
The time evolution of $r_\mathrm{g}$ for individual trajectories can be found in Figure~S6.
At the beginning of each trajectory, the dielectron is noticeably more delocalized due to the initially oversized cavity inherited from the I$^-$ placeholder used during the preparation phase.
As the simulation progresses, the cavity reorganizes and the electron becomes more localized, reaching an equilibrated gyration radius value.
This happens within less than the 1~ps that we discard from our analysis for equilibration.
The resulting average values of $r_\mathrm{g}$ are approximately 2.5~\AA\ for systems containing cations and 2.3~\AA\ for the system without cations.

The shape descriptors, namely asphericity, acylindricity, and relative shape anisotropy $\kappa^2$, were evaluated from the gyration tensor in the same manner as $r_\mathrm{g}$.
Their distributions for all systems are shown in Figures~\ref{fig:gyration-descriptors} and \ref{fig:2D-r-k}.
Asphericity quantifies the overall deviation of the electron density from spherical symmetry, acylindricity describes the deviation from cylindrical symmetry through the difference between the two smaller principal components of the gyration tensor, and $\kappa^2$ provides a dimensionless overall measure of anisotropy ranging from 0 for a perfectly spherical distribution to 1 for a highly elongated shape.
The dielectron in the presence of cations is found to be slightly less spherical compared to the system without cations, as indicated by asphericity, while the acylindricity and $\kappa^2$ remain essentially unchanged across all systems.
To further analyze the relationship between electron size and shape, two-dimensional histograms of $r_\mathrm{g}$ versus $\kappa^2$ were constructed for each system and are presented in Figure~\ref{fig:2D-r-k}.
Neither strong correlation between gyration radius and relative shape anisotropy nor systematic differences between the studied systems are apparent from these distributions, though we will revisit this relationship later in the discussion of reactive trajectories.

\subsection{Dielectron--cation interactions}

\begin{figure}[]
    \centering
    \includegraphics{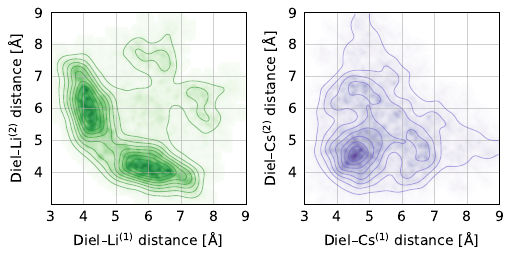}
    \caption{Two-dimensional probability density of distances between the dielectron center and individual metal cations for Li$^+$ (left) and Cs$^+$ (right).}
    \label{fig:el-M-dists}
\end{figure}

The explicit inclusion of metal cations allows us to directly investigate dielectron--cation interactions, their preferred spatial arrangements, and the influence of cations on dielectron properties.
First, we look at the correlation of the distances between the dielectron and the two metal cations, shown in Figure~\ref{fig:el-M-dists}.
The difference between the Li$^+$ and Cs$^+$ systems is immediately apparent, with the two distributions appearing almost complementary.
An additional perspective on the distance correlation is provided in Figure~S7, where the metal--metal distance is introduced.
For the lithium system, most of the configurations typically have one Li$^+$ ion close to the dielectron, with the second Li$^+$ ion much further away.
The highest probability corresponds approximately to one Li$^+$ at 4~\AA{} and the second at 6~\AA{} from the dielectron.
In contrast, the cesium system exhibits a preferred configuration where both Cs$^+$ ions are located at similar distances of approximately 4.5~\AA{} from the dielectron, while asymmetric configurations are less probable.
Representative snapshots corresponding to these probability maxima are shown in Figure~\ref{fig:snapshots}.
Interestingly, the Li$^+$ distribution is slightly tilted, indicating that bringing one lithium ion closer to the dielectron pushes the second ion toward almost the largest distances accessible within the simulation box.

These observations are consistent with the dielectron--cation RDFs (top right panel of Figure~\ref{fig:rdfs-all}), where the first RDF maxima appear at approximately 4~\AA{} for Li$^+$ and 4.5~\AA{} for Cs$^+$.
At the same time, despite the shifted peak positions, the onset of the first RDF peak occurs at nearly the same distance for both systems.
The contact ion pair configuration observed for Li$^+$, visible in the RDF as a small peak around 2.2~\AA{}, is not apparent in the two-dimensional distributions due to the limited number of configurations, however they would correspond to distances of 2.2~\AA{} and 7.0~\AA{} from the dielectron.

Next, we examine the effect of dielectron--metal distances on the dielectron shape descriptors to see whether their distributions change as the dielectron approaches the cation.
Surprisingly, only very minor effects are observed.
For the Cs$^+$ system, a slight increase in gyration radius is observed at shorter dielectron--cation distances.
However, the asphericity and acylindricity distributions remain essentially unchanged irrespective of whether the dielectron is close to or far from the cation.
For the Li$^+$ system, all three analyzed shape descriptors remain nearly identical across the entire range of dielectron--cation distances.
An alternative representation of the same data is provided in Figure~S8, where the distributions of the shape descriptors are normalized separately for each dielectron--cation distance.

\begin{figure}[]
    \centering
    \includegraphics{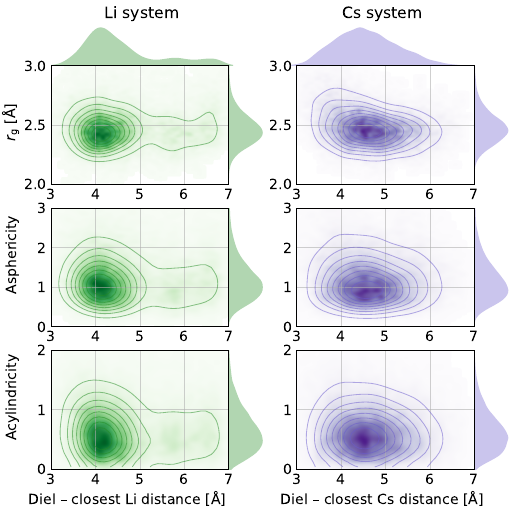}
    \caption{Correlation between dielectron shape descriptors ($r_{\mathrm{g}}$, asphericity, and acylindricity) and the distance to the nearest metal cation for Li$^+$ (left) and Cs$^+$ (right).
    }
    \label{fig:desc-dist-correlation}
\end{figure}

\subsection{Vibrational properties from structure}

\begin{figure}[htbp]
    \centering
    \includegraphics{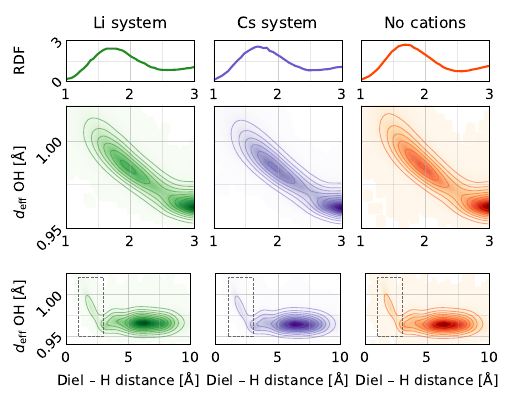}
    \caption{Top panels: dielectron–hydrogen radial distribution functions for each system.
    Middle panels: effective O–H bond length as a function of dielectron–H distance, shown for the 1--3~\AA\ region.
    Bottom panels: effective O–H bond length as a function of dielectron–H distance over the full range, with the zoomed region indicated by a dashed gray line.
    }
    \label{fig:OH bond length}
\end{figure}

To characterize the vibrational properties of the simulated systems, we build on recent work~\cite{boyer2019beyond, santis2025extending1, santis2025extending2} that establishes a broadly-applicable relationship between instantaneous vibrational frequencies and instantaneous equilibrium covalent bond lengths in aqueous hydrogen bonds at a finite temperature.
This approach allows us to predict vibrational properties from structures without the need to perform actual (localized) vibration analysis of the trajectories.
Following the protocol introduced in Ref.~\citenum{boyer2019beyond}, we filter out all frequency components above 1000~cm$^{-1}$, obtaining an effective O--H bond length where the slower structural response to the local environment is retained.

To localize shifts in vibrational frequencies, we correlate the effective O--H bond lengths $d_{\mathrm{eff}}(\mathrm{OH})$ with the distance of their hydrogen atom to the hydrated dielectron.
The resulting distributions in Figure~\ref{fig:OH bond length} reveal a systematic trend.
For ``normal'' O--H bonds far from the dielectron, $d_{\mathrm{eff}}(\mathrm{OH})$ is centered around $\sim$0.96~\AA, consistent with the bulk liquid water value.
In contrast, when the O--H bond gets closer to dielectron, its effective bond length increases, reaching values up to $\sim$1.02~\AA{} for hydrogens located slightly above 1~\AA{} from the center of the dielectron.

This distance range of maximum elongation (1--2~\AA{}) corresponds to hydrogens in the first solvation shell of the dielectron, as also evident from the diel--H radial distribution functions shown for reference in the top panel of Figure~\ref{fig:OH bond length}.
Importantly, the elongation is observed only for O--H bonds oriented toward the dielectron, whereas the opposite O--H bonds within the same water molecule, pointing away from the electron, retain their bulk-like bond length.
This elongation corresponds to an increase of about 5\%, which is a substantial elongation for a covalent bond.
The same behavior is observed in systems with or without cations, with no noticeable differences between the data.

\subsection{Reactivity}

Out of our 30 trajectories without cations, two exhibited a complete reaction leading to molecular hydrogen formation at very early stages of the simulation (after 200~fs and 400~fs from the start of the production run).
In the first step of that reaction, a proton from an O--H bond in the first solvation shell of the dielectron is transferred toward the center of the excess electron density, forming a hydride intermediate and a hydroxide anion.
This hydride species rapidly accepts a second proton transfer from another water molecule in the solvation shell, yielding molecular hydrogen and a second hydroxide anion.
This early in the trajectories, the transition from the iodide used to create the cavity to the dielectron is not fully equilibrated and thus these reactions are likely an artifact of the simulation protocol.
As such, we do not analyze them in more detail here.

In two additional trajectories, only the initial proton transfer event was observed, resulting the formation of hydride that was stable until the end of the trajectory.
These events occurred later in the simulation (at approximately 1.5~ps and 3.5~ps), indicating that the system was already equilibrated.
Once the hydride was formed, no further reaction was observed over the remaining simulation time (6.5~ps and 2.5~ps).
In the following, we focus on the analysis of the these two trajectories and the hydride product, which represents the first step of the hydrogen evolution reaction.

\begin{figure}[t]
    \centering
    \includegraphics{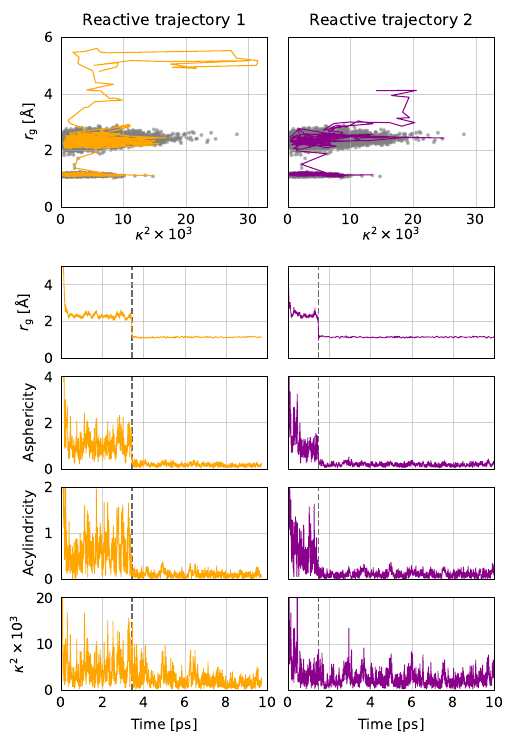}
    \caption{Top panels: two-dimensional distributions of gyration radius $r_{\mathrm{g}}$ and relative shape anisotropy $\kappa^2$ for the two reactive trajectories leading to hydride formation. Bottom panels: time evolution of gyration parameters ($r_{\mathrm{g}}$, asphericity, acylindricity, and $\kappa^2$) for the same reactive trajectories.}
    \label{fig:hydrides-all}
\end{figure}

First, we follow the shape descriptors of the solute electron density along the reactive trajectories.
In the top panel of each column of Figure~\ref{fig:hydrides-all}, we correlate the gyration radius and the relative shape anisotropy and reference them against the distribution from all trajectories.
In the remaining panels, we show the time evolution of the gyration radius, asphericity, acylindricity, and relative shape anisotropy.
At early times, the dielectron is far from its equilibrium values of all descriptors due to the initial conditions, which is typical for both reactive and non-reactive trajectories.
After equilibration, the system remains for at least 1~ps in a well-defined region of $r_{\mathrm{g}}$ and $\kappa^2$ values before the proton transfer occurs.
The formation of the hydride is marked by a sharp decrease of the gyration radius from approximately 2.3~\AA\ to about 1.2~\AA, accompanied by a large decrease of asphericity and acylindricity to almost zero, and a slight reduction in anisotropy.
There is a marked decrease of the fluctuations of all the quantities with the exception of anisotropy, which decreases less, and more gradually.
This behavior reflects the transition from a diffuse, highly distorted dielectron density to a compact and nearly spherical hydride state, resembling an s-type orbital.

\begin{figure}[t]
    \centering
    \includegraphics{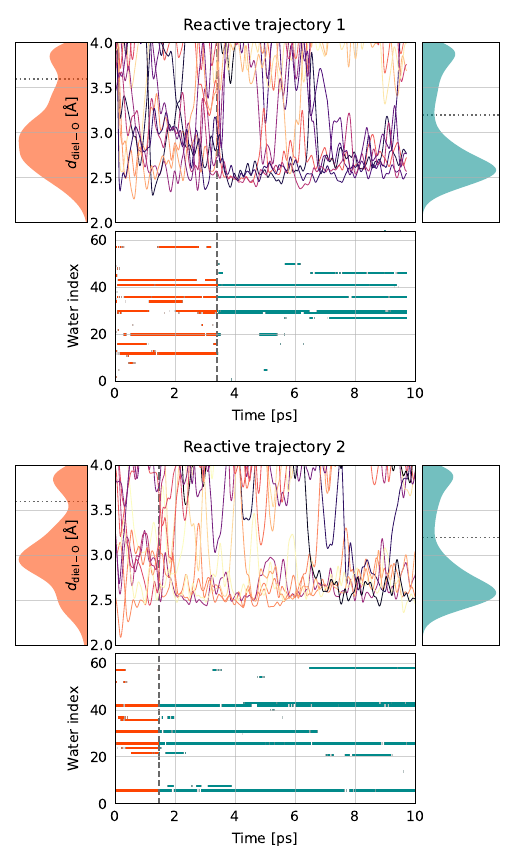}
    \caption{Solvation dynamics of two reactive trajectories leading to hydride formation.
    Top panels: time evolution of dielectron–oxygen distances within 4~\AA\ from the dielectron center (smoothed using a Gaussian filter with \mbox{$\sigma = 10$}), with the dashed gray line indicating the time of hydride formation.
    Side histograms show the distributions of dielectron–oxygen distances before (left) and after (right) the reaction.
    Bottom panels: indices of water molecules within 3.6~\AA\ before the reaction and within 3.2~\AA\ after the reaction, illustrating the composition of the first solvation shell for the dielectron and the hydride.
    }
    \label{fig:hydride-solvation}
\end{figure}

\begin{figure}[]
    \centering
    \includegraphics{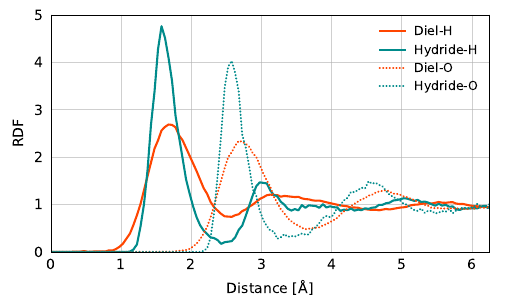}
    \caption{Hydride--water radial distribution functions of reactive trajectories (cyan) compared with dielectron--water RDFs shown earlier in Figure~\ref{fig:rdfs-all}.}
    \label{fig:hydride-RDF}
\end{figure}

We next examine the solvation dynamics and the changes in the solvation shell accompanying the transformation of the dielectron into a hydride.
The time evolution of the solvation structure is summarized in Figure~\ref{fig:hydride-solvation}.
For each trajectory, the top panel shows the distances between the dielectron and oxygen atoms within 4~\AA\, with the time of the hydride formation indicated by a vertical dashed gray line.
The side panels show the distributions of these distances before (left) and after (right) the proton transfer.
A clear shift in solvation structure from the dielectron to the hydride can be observed.
The first solvation shell of the dielectron is relatively broad and ends at approximately 3.6~\AA\ from its center.
In contrast, the hydride exhibits a more compact solvation structure, with a sharper first maximum and a deeper first minimum, located at around 3.2~\AA.
These characteristic distances were used to define the first solvation shell in the lower panels in Figure~\ref{fig:hydride-solvation}, where the time evolution of solvent indices within 3.6~\AA\ (before the reaction) and 3.2~\AA\ (after the reaction) is shown.
This representation provides insight into the instantaneous size and composition of the first solvation shell, complementing the information obtained from integrated coordination numbers.
On the investigated timescales, both the dielectron and the hydride maintain relatively stable solvation shells, with water exchange events occurring only rarely and no substantial large-scale rearrangement of surrounding molecules.
The dielectron is typically coordinated by 6--7 water molecules, whereas the hydride is stabilized by a smaller number of 4--5 water molecules, reflecting a significant reorganization of the solvation environment upon proton transfer.
For a direct comparison, in Figure~\ref{fig:hydride-RDF}, we reference the hydride--water RDFs from these trajectories against the dielectron--water RDFs shown earlier in Figure~\ref{fig:rdfs-all}.
This clearly shows how the hydride solvation shell becomes smaller and more structured.

\section{Discussion and conclusions}

In this work, we performed AIMD simulations of hydrated dielectrons in bulk water either with no counter ions included or in the presence of Li$^+$ or Cs$^+$ cations.
We focus on how the presence of alkali metal cations influences the structure, vibrational properties, and reactivity of hydrated dielectrons, and on description of dielectron--cation interactions in solution.
Before discussing these results in detail, it is useful to compare our simulations with the two previous AIMD studies of hydrated dielectrons in bulk water~\cite{reaction_borrelli,reactions_gao}, since several methodological differences may influence the obtained structural properties.

In our work, the systems were prepared using an iodide anion to form the cavity, before exchanging it for a dielectron without any change of water geometry.
This replacement is then followed by a roughly 1~ps equilibration period at the beginning of our production trajectories, employing the CSVR thermostat with a 500~fs time constant.
In contrast, previous studies~\cite{reaction_borrelli,reactions_gao} started from an equilibrated hydrated electron and introduced the second electron vertically at the beginning of the production run, where CSVR or Nose--Hoover chains thermostats with unreported settings were used.
The size of the NVT simulation cell is also determined differently here.
While the previous studies assumed a zero molar volume of the dielectron, we accounted for the estimated molar volumes of the solutes, resulting in a slightly expanded cell.
Finally, differences may also arise from the employed electronic structure methods and basis sets, with the previous studies having used PBE0-D3 with the default 25\% of HF exchange and ``a triple-$\zeta$'' basis set~\cite{reaction_borrelli} or the PBE0-rVV10 functional with 40\% HF exchange and a TZVP basis set~\cite{reactions_gao}.

\subsection{Structure of hydrated dielectrons}

Despite the methodological differences mentioned above, the radial distribution functions obtained in the present work are qualitatively very similar to those reported previously~\cite{reaction_borrelli,reactions_gao}.
Although direct comparison is not possible since coordination numbers were not reported in the previous studies, the good agreement between the RDFs suggests a consistent picture.
The present coordination shell consists of 6--7 water molecules around the dielectron, which is slightly larger than that reported for the hydrated monoelectron, where 4--5 water molecules are typically found in the first solvation shell~\cite{park2022understanding,kar2025nature}.

The most notable structural influence of metal cations observed in the present work is the size of the excess dielectron.
Namely, the average gyration radius of the hydrated dielectron increases from approximately 2.3~\AA{} in the system without cations to 2.5~\AA{} in the presence of Li$^+$ or Cs$^+$, whereas the remaining shape descriptors change only slightly.
Interestingly, this difference in $r_\mathrm{g}$ between systems with and without cations is not directly reflected in the size of the solvent cavity, which appears comparable based on the radial distribution functions (Figure~\ref{fig:rdfs-all}).
These observations suggest that although the cavity size remains similar, the spatial extent of the electron density differs, raising the question of how the presence of cations modifies the electronic structure without significantly altering the underlying solvent structure.

Comparing the gyration radius with previous calculations reveals additional issues.
Both earlier AIMD studies reported substantially larger dielectron gyration radii of 2.68~\AA{} and 2.91~\AA{} for the corresponding system without cations~\cite{reaction_borrelli,reactions_gao}, compared to the value of approximately 2.3~\AA{} obtained here.
Experimental estimates of electron size are available only for hydrated monoelectrons, for which values between 2.4 and 2.7~\AA{} have been inferred from spectroscopic measurements~\cite{kevan1981electron,bartels2001moment,bartels2005pulse,coe2008photoelectron}.
However, no direct experimental estimates of hydrated dielectron size have been reported so far, making it difficult to assess which of the above computations is more accurate.

\subsection{Dielectron–cation interactions}

Having discussed the structural properties of hydrated dielectrons, we now turn to the dielectron--cation interactions themselves.
Since the explicit inclusion of metal cations is one of the main novelties of the present work, it is natural to ask how the cations are arranged around the dielectron and whether their presence affects the excess dielectrons properties.
Given that ion pairing is a well-established phenomenon in solution chemistry, dielectron--cation association and partial localization effects might also be expected in these systems.

The dielectron--cation interactions differ substantially between Li$^+$ and Cs$^+$ containing systems.
Li$^+$ preferentially forms asymmetric configurations with one cation close to the dielectron and the second farther away, whereas Cs$^+$ favors more symmetric structures with both cations located at similar distances from the dielectron.
This indicates that the nature of the alkali metal cation plays an important role in determining the microscopic organization of excess-electron systems, consistent with previously observed cation-dependent behavior of alkali-metal aqueous solutions~\cite{vitek2025dynamics}.
Despite this substantial cationic effect, the dielectron shape descriptors depend only weakly on the dielectron--cation distance.

\subsection{Vibrational properties from structure}

The presence of excess electrons strongly perturbs the surrounding water molecules.
Earlier resonance Raman measurements of the bulk hydrated electron reported substantial red-shifts of the O--H stretching band, of the order of 200~cm$^{-1}$~\cite{tauber2003structure,tauber2001fluorescence,tauber2002resonance,mizuno2003picosecond}, which were interpreted as evidence for weakened O--H bonds in water molecules coupled to the electron.
Such red-shifts are not unique to hydrated electrons, but rather represent a general hallmark of hydrogen bonding observed across a wide range of systems and chemical environments~\cite{bhatta2014brief}.
The origin of these spectral features for the hydrated electron was investigated theoretically, where similar red-shifts were qualitatively reproduced in infrared spectra of anionic water clusters~\cite{herbert2006charge} and in resonance Raman spectra of the bulk hydrated electron~\cite{Herbert_Raman}.
These spectral changes were attributed to significant charge transfer and consequent partial occupation of antibonding $\sigma^*_{\mathrm{OH}}$ orbitals.

Our results provide direct structural evidence that the hydrated dielectron selectively elongates nearby O--H bonds.
Within the framework of instantaneous beyond-Badger-type correlations~\cite{boyer2019beyond}, this elongation implies a decrease in the O--H force constant and a corresponding red-shift of the stretching frequency, in agreement with experimentally reported frequency shifts.
We find that the effect is highly localized and originates from O--H bonds belonging to water molecules in the first solvation shell and oriented toward the dielectron.
These bonds exhibit a substantial elongation compared to bulk water, whereas O--H bonds pointing away from the dielectron remain essentially unaffected.
The present results therefore provide direct structural support for the interpretation of the resonance Raman experiments and establish a microscopic link between the local solvation structure of the dielectron and its vibrational signatures.

\subsection{Reactivity}

Both in the present work and in previous AIMD studies, reactive events were observed to occur as two sequential proton-transfer steps via a hydride intermediate, in agreement with the mechanism proposed by Bartels~\cite{reaction_bartels}.
However, the number and rates of the reactive events seem to be very different.
In the previous studies, most of the dielectrons reacted within 130--2000~fs from the start of simulation, with the second proton transfer always following the first one in a 30--200~fs time window.
The simulations were terminated if no reaction occurred within a 2~ps limit.
In contrast, out of our 30 trajectories of lengths of 5--10~ps, only two reacted completely at very early stages of the simulation (200~fs and 400~fs) before proper equilibration.
Taken together, this may suggest that the high reactivity reported previously is at least partially due to incomplete equilibration of the initial configurations.
This raises the possibility that the hydrogen-evolution events reported so far, including those observed in the present work, originate from dielectron configurations that are not fully equilibrated.
If this is indeed the case, neither the previous AIMD studies nor the present work may have realistically captured the process of hydrogen evolution from the dielectron, highlighting the need for future studies aimed at capturing hydrogen evolution from a fully equilibrated hydrated dielectron.
At the same time, the present work provides an improved description of the hydride intermediate compared to previous studies.
In our simulations, two additional trajectories formed hydride intermediates at substantially later times (1.5~ps and 3.5~ps), with no additional proton transfer observed for the remainder of the trajectories.
Although the experimental lifetime of the hydride remains unknown due to its high reactivity, it was observed as a reactive intermediate in related radiation-induced reactions~\cite{kimmel1994low, bernas1977existence, laverne2000new, marin2007recombination}.

The above mentioned studies also suggested that the hydride solvation structure closely resembles that of the dielectron, reporting a very similar RDF for both species~\cite{reaction_borrelli}.
However, these observations likely primarily reflect the lack of solvent reorganization due to the extremely short hydride lifetime.
In contrast, the longer-lived hydride observed in our work clearly exhibits a distinctly different solvation structure from that of the dielectron, being stabilized by only 4--5 water molecules.
Similarly, the size of the hydride species, calculated from its electron density, is notably smaller in our work ($\sim$1.2~\AA) compared to the referenced articles ($\sim$1.7~\AA).
The hydride RDF thus shows substantial changes relative to the dielectron, indicating a more structured solvation environment and a smaller solvent cavity.
It is also important to note that proton-transfer leading to hydride formation does not pass through a distinct transition-state-like region characterized by increased anisotropy and intermediate values of $r_{\mathrm{g}}$.
Instead, the transition preserves values of anisotropy, however, with a sharp decrease in the gyration radius.
This behavior differs from proton transfer to the hydrated monoelectron, where a highly anisotropic transition state was observed~\cite{marsalek2010hydrogen}.

As noted earlier, no proton transfer events were observed in any of our simulations containing metal cations, with neither hydride formation nor molecular hydrogen production occurring within the investigated timescales.
As a further check, we performed additional simulations of the system with lithium cations using an independent computational package (VASP)~\cite{vasp1, vasp2, vasp3, vasp4}, as well as using the revPBE0-D3 functional (25\% HF exchange) using both CP2K and VASP.
Importantly, none of these additional trajectories exhibited reactive behavior either (see details Table~S1).
This observation may suggest that the presence of metal cations suppresses the reactivity of hydrated dielectrons on the investigated timescales.
At the same time, it remains unclear whether this behavior reflects a genuine physical effect of the cations or or is due to technical differences between the simulations of the three systems.
Note in this context that comparison between CP2K and VASP simulations revealed additional differences that are beyond the scope of the present work, and will be addressed in a forthcoming methodological study.

\section*{Supporting Information}

The Supporting Information contains individual RDFs and integrated coordination numbers for all studied systems, together with separate water--water RDFs for molecules closest to the dielectron and for the remaining solvent.
Time evolution of the gyration radius for all individual trajectories is also provided.
For the dielectron--cation interactions, correlation of dielectron--metal distances and metal--metal distances is presented, as well as correlations between dielectron shape descriptors and the normalized distance to the nearest metal cation.
Furthermore, examples of two-dimensional distributions of gyration radius and relative shape anisotropy for the different systems and representative examples of dielectron solvation dynamics are shown.
Finally, a summary table of all simulated systems and the observed reactive events is included.

\section*{Acknowledgment}

T.N. acknowledges support from University of Chemistry and Technology Prague where she is enrolled as a PhD. student and from the IMPRS  for Quantum Dynamics and Control.
The authors also acknowledges Dr. Jinggang Lan for fruitful discussions on the reactivity of solvated dielectrons.
P.J. acknowledges support from the Czech Science Foundation via grant no. 24-10982S. 

\section*{References}
\bibliography{paper}

\end{document}